\documentclass[intlimits,twoside,a4paper]{article}
\usepackage{graphicx}
\usepackage{amsmath,amssymb}

\usepackage[T2A]{fontenc}
\usepackage[cp1251]{inputenc}

\usepackage{cmpj2}

\issue{2014}{17}{1}{13702}
\doinumber{10.5488/CMP.17.13702}

\title[Effect of magnetic field on electron spectrum in spherical
nano-structures]{Effect of magnetic field on electron spectrum in spherical
nano-structures}

\author[V. Holovatsky, O. Voitsekhivska, I. Bernik]
{V. Holovatsky\thanks{E-mail: ktf@chnu.edu.ua}\,, O. Voitsekhivska, I. Bernik}%
\address{Chernivtsi National University, 2~Kotsiubynsky St.,
58012 Chernivtsi, Ukraine}

\date{Received September 28, 2013, in final form November 3, 2013}
\authorcopyright{V. Holovatsky, O. Voitsekhivska, I. Bernik, 2014}

\begin{document}

\maketitle

\begin{abstract}
The effect of a magnetic field on the energy spectrum and on the wave
functions of an electron in spherical nano-structures such as single
quantum dot and spherical layer is investigated. It is shown that
the magnetic field removes the spectrum degeneration with
respect to the magnetic quantum number. An increasing magnetic field
induction entails a monotonous character of electron energy for
the states with $m \geqslant 0$ and a non-monotonous one for the states
with $m<0$. The electron wave functions of the ground state and several
excited states are studied considering the effect of the
magnetic field. It is shown that $1s$ and $1p$ states are degenerated
in the spherical layer driven by a strong magnetic field. In the
limit case, a series of size-quantized levels produce the Landau
levels which are typical of bulk crystals.

\end{abstract}

\section{Introduction}

The multilayered spherical nano-structures consisting of a core
and a few spherical shells attract a particular attention
of scientists. These structures are grown using the chemical
colloidal method on the basis of CdS, CdSe, ZnS, ZnSe, HgS and
other semiconductor materials \cite{1,2,3}. Multilayered nano-structures
have wide prospects for being  utilized in medicine and electronics,
for instance for the purpose of fabricating efficient biosensors and fluorescent labels \cite{4,5}. Multilayered spherical
nano-structure containing two quantum wells formed by the core
and spherical layer is called quantum-dot-quantum-well (QDQW).
Such structures can be the basis for modern highly efficient white
light sources \cite{6}. In QDQW, a quasi-particle can be located in
one of the quantum wells. The peculiarities of electron and hole
location in multilayered spherical nano-structures have been
theoretically studied in \cite{7,8,9,10}. The impurities, external
electrical and magnetic fields produce an effect on the location of
quasi-particles in quantum wells \cite{11,12,13,14,15,16,17,18,19} and consequently effect the optical
properties of the structures. The on-center Coulomb impurity does
not violate the spherical symmetry of the system and thus the
Schr\"odinger equation for an electron or hole is solved exactly
\cite{11,12,13}. When the external fields are present, the spherical
symmetry is violated and the calculation of an energy spectrum becomes
more complicated due to the fitting conditions at the interfaces
\cite{14,15,16}. Therefore, in the majority of theoretical studies, the
investigation of ground state energies of quasi-particles are
performed within the variational method in the frames of the model
of infinitely deep potential well \cite{17,18}. The effect of
a magnetic field on the energies and on wave functions of
the excited states of quasi-particles is still insufficiently investigated
for a spherical quantum dot. The analogous problem for QDQW has
not been studied at all.

In the case of high potential barrier of QDQW, the electron in low
states does not penetrate through the interfaces of the system and
quantum wells become decoupled \cite{20}. The effect of a magnetic
field on the states of electron located in the core or in the spherical
layer of QDQW is different and can be investigated considering
these two potential wells independently. Therefore, in this paper
we study the magnetic field effect on the energy spectrum and  on the wave
functions of an electron located
 in the quantum dot (QD) and in the spherical layer
(SL), assuming that the potential barriers are infinite.

\section{Schr\"odinger equation for the electron
in spherical nano-structures driven by magnetic field}

We consider the spherical QD with the radius $r_0$ and the spherical layer
with the inner and outer radii $r_1$ and $r_2$, respectively, having
impenetrable boundaries. The coordinate system is taken in such a
way that its origin is in the center of the structure and Oz axis
coincides with the direction of the magnetic field induction.

The potentials of size quantization for the electron are as follows:
\begin{equation} \label{EQ.5a}
U^{(0)} (r)=\left\{
\begin{array}{ll}
0, & \hbox{$r\leqslant r_{0}$}\,,  \\
\infty, & \hbox{$r>r_{0}$} \,,
\end{array}\right.
\end{equation}
\begin{equation} \label{EQ.5b}
 U^{(1)} (r)=\left\{
\begin{array}{ll}
0, & \hbox{$r_{1} \leqslant r\leqslant r_{2}$}\,,  \\
\infty, & \hbox{$r<r_{1}$, \  $r>r_{2}$}\,.
\end{array}\right.
\end{equation}
Here, $U^{(0)}$  is the potential energy of an electron in
spherical QD and  $U^{(1)}$~--- in SL having impenetrable boundaries.

Schr\"odinger equations for the electron in these systems in
a magnetic field are as follows:
\begin{equation} \label{EQ.1}
\left[\frac{1}{2 \mu}\left(\vec{p}-\frac{e}{c} \vec{A}\right)^2 +U^{(0,1)}(r)\,\right]
\psi^{(0,1)} \left(\vec{r}\right)=E^{(0,1)}\, \psi^{(0,1)}
\left(\vec{r}\right).
\end{equation}

When $\vec{A}=[\vec{r}\times \vec{B}]/2$, the Hamiltonians become
\begin{equation} \label{EQ.4}
H^{\left(0,1\right)} =-\frac{\hbar ^{2} }{2 \mu } \Delta
+\frac{eB}{2\, c\, \mu } L_{z} +\frac{e^{2} B^{2} r^{2} \sin ^{2}
\theta }{8\, c^{2} \, \mu } +U^{\left(0,1\right)} (r),
\end{equation}
where $L_{z} =-\ri\hbar \, \partial /\partial \varphi $.

Using the dimensionless magnitudes:
 $R^{*} =e^{2}
/(2\, \varepsilon \, a^{*} )$~--- effective Rydberg energy, $a^{*}
=\hbar ^{2} \varepsilon /(\mu  e^{2} )$~--- effective Bohr radius
and parameter  $\eta = \hbar \omega_{\mathrm{c}}/(2 R^{*} )$, where $\omega_{\mathrm{c}}
= B e/(\mu c)$~--- cyclotron frequency, the Hamiltonian (\ref{EQ.4}) is
transformed into
\begin{equation} \label{EQ.6}
H^{(0,1)} =-\Delta +\eta L_{z} \, +\frac{1}{4}(\eta \, r\sin \theta )^{2} +U^{(0,1)} (r).
\end{equation}
When $\eta =0$, the Schr\"odinger equation with Hamiltonian (\ref{EQ.6})
has the exact solutions
\begin{equation} \label{EQ.7}
\Phi _{n\, l\, m}^{^{(0,1)} } (r,\theta ,\varphi )=R_{n\, l}^{^{(0,1)} } (r)Y_{lm} (\theta ,\varphi ),
\end{equation}
where
$R_{nl}^{(0)} (r)\, =A_{nl}^{(0)} j_{l} (\chi _{nl} r/r_{0} )$,
$R_{nl}^{(1)} (r)\, =A_{nl}^{(1)} j_{l} (k_{nl}r)+B_{nl}^{(1)} n_{l} (k_{nl} r)$,
$A_{nl}^{(0)} =\sqrt{2} /\big[r_{0} ^{3/2} j_{l+1} (\chi _{nl} )\big]$,
$\,\,\, B_{nl}^{(1)} =-A_{nl}^{(1)} \, j_{l} (k_{nl} r_{1} )/n_{l}
(k_{nl} r_{1} )$,
$j_l(z)$, $n_l(z)$ are Bessel spherical functions of the first
and the second kind, respectively,
$\chi_{nl} $ are the roots of Bessel spherical function
[$j_l (\chi _{nl} )=0$], the values $k_{nl} $ are fixed by the
condition $R_{nl}^{(1)} (r_{2} )=0$ and the coefficients
$A_{nl}^{(1)} $ are fixed by the normality condition.

The square term (with respect to the magnetic field) in the
Hamiltonian (\ref{EQ.6}) rapidly increases its contribution into the
complete energy. Thus, it is impossible to use the perturbation
method. In order to obtain the ground state energy using the  variational
method it is necessary to define the approximated wave function.
In \cite{21} it is written as follows:
\begin{equation} \label{EQ.8}
\psi _{100}^{^{(0,1)} } (\vec{r})=\re^{-g(\vec{r})} \Phi
_{100}^{^{(0,1)} } (\vec{r}),
\end{equation}
where the function $g(\vec{r})$ is to ensure the compensation of
a quadratic term  after substitution of (\ref{EQ.8}) into (\ref{EQ.6}). The minimum
condition for
\begin{equation} \label{EQ.9}
F(\eta)=\frac{1}{4}(\eta \, r\, \sin \theta )^{2} -\left|\vec {\nabla}
g(\vec{r})\right|^{2}+\Delta g(\vec{r})
\end{equation}
at  $\left. g(\vec{r})\right|_{\eta \to 0} =0$ fixes
$g(\vec{r})=\eta r^2 \sin^2 \theta /4$. Herein, $F(\eta)$ linearly
depends on $\eta$, thus, the variational function of the electron
ground state can be written as follows:
\begin{equation} \label{EQ.10}
\psi _{100}^{(0,1)} (\vec{r})=C\, \exp \left(-\eta r^2 \sin^2
\theta /4\right)\Phi _{100}^{(0,1)} (\vec{r})\exp (\lambda \, r),
\end{equation}
where $C$ is the normality constant and   $\lambda$ is the
variational parameter. The form of the variational wave function
is confirmed by physical considerations. The magnetic field
directed along Oz axis deforms the wave function compressing it in
perpendicular direction. This fact is represented by an angular
dependence of the function (\ref{EQ.10}). Variational parameter $\lambda$ and
the energy of ground state are obtained from the minimum of
the whole energy
\begin{equation} \label{EQ.11}
E_{100}^{(0,1)} =\mathop{\min }\limits_{\lambda } {\left\langle
\psi _{100}^{(0,1)} (\vec{r}) \right|} H^{(0,1)}{\left| \psi
_{100}^{(0,1)*} (\vec{r}) \right\rangle} .
\end{equation}

In order to study the excited states, the orthonormality condition
for the wave functions should be fulfilled, which makes the problem
rather complicated. Therefore, we are going to use another
method to solve it. We expand the wave function using a
complete set of eigenfunctions of the electron in a spherical
nano-structure without the magnetic field obtained as the exact
solutions of Schr\"odinger equation \cite{15}. When the magnetic field
is applied, the spherical symmetry is violated and the orbital
quantum number becomes inconvenient. The new states
characterized by a magnetic quantum number $m$ are presented as a
linear combination of the states  $\Phi _{n\, lm}^{(0,1)}
(\vec{r})$
\begin{equation} \label{EQ.12}
\psi _{j\, m}^{(0,1)} (\vec{r})=\sum _{n}\sum _{l}c_{n\, l} \Phi _{n\, lm}^{(0,1)} (\vec{r})  .
\end{equation}

Substituting (\ref{EQ.12}) into Schr\"odinger equation with Hamiltonian
(\ref{EQ.6}), we obtain a secular equation for the electron energy
spectrum
\begin{equation} \label{EQ.13}
\left|H_{n\, l,n'l'}^{(0,1)} -E_{jm}^{(0,1)} \delta _{n,n'} \delta _{l,l'} \right|=0,
\end{equation}
where the matrix elements  $H_{n'\, l',n\, l}^{(0,1)} $ have the
form
\begin{equation}\label{EQ.14}
H_{n'\, l',n\, l}^{(0,1)} =\left(E_{nl}^{0(0,1)}+m\,
\eta\right) \delta _{n'n} \delta _{l'l}+ \frac{\eta ^{2}}{2}
\left\{\alpha _{l,m} \delta _{l',l+2} +\beta _{l,m} \delta _{l',l}
+\gamma _{l,m} \delta _{l',l-2} \right\}\, I_{n'\, l',n\,
l}^{(0,1)} \,,
\end{equation}
\begin{align*}
 I_{n'l',nl}^{(0)} &=A_{n'l'}^{(0)} A_{nl}^{(0)} \int
_{0}^{r_{0} }r^{4} j_{\,l'}
(\chi_{n'l'} \, r/r_0)j_{\,l} (\chi_{nl} \, r/r_0)\rd r, \\
 I_{n'l',nl}^{(1)} &=\int _{r_{1} }^{r_{2} }r^{4}
\left[A_{n'l'} j_{l'} (k_{n'l'} r)+B_{n'l'} n_{l'} (k_{n'l'} r)\right]\left[A_{nl}
j_{l} (k_{nl} r)+ B_{nl}
n_{l} (k_{nl} r)\right]\rd r, \\
 \alpha _{l,m} &=-\sqrt{
\frac{[(l+2)^2-m^2][(l+1)^2-m^2]}{(2l+5)(2l+3)^{2}
(2l+1)} } \,, \\
 \beta _{l,m} &=1-\frac{(l+1)^2- m^2}{(2l+1)(2l+3)}
-\frac{l^2-m^2}{4l^2-1}, \qquad
\gamma _{l,m} =-\sqrt{\frac{(l^2-m^2)[(l-1)^2-m^2]}{(2l+1)(2l-1)^{2} (2l-3)} }\,.
\end{align*}

Using the eigenvalues and eigenvectors of the matrix
\begin{equation} \label{EQ.15}
F_{n\, l,n'l'}^{(0,1)} =H_{n\, l,n'l'}^{(0,1)} -E_{jm}^{(0,1)} \delta _{n,n'} \delta _{l,l'}\, ,
\end{equation}
we obtain the energy spectrum and wave functions of the electron
in a spherical nano-structure driven by the  magnetic field.

\section{Analysis of the results}

The computer calculations were performed using the physical
parameters of CdSe semiconductor material: electron effective mass
$\mu =0.13\, m_{e} $ ($m_e$~--- the mass of pure electron),
dielectric constant $\varepsilon=10.6$.

Expanding the wave functions (\ref{EQ.14}) we took into account a sufficient number of terms, provided that the sum of squares
of expansion coefficients was equal to a unit with the accuracy not
less than 0.01\%. In table~\ref{tabl:table1} the expansion coefficients for
the lowest states at $B=40$~T are presented. Here one can see that
the required accuracy is provided by 6 major terms. For
convenience, we use the same quantum numbers characterizing the
states of an electron in a nano-structure driven by a magnetic field as
the ones without the field.

\begin{table}[!t]
\centering
 \caption{Expansion coefficients of electron wave functions $\psi_{100}^{^{(0)} }$, $\psi _{110}^{^{(0)} }$, $\psi_{11-1}^{^{(0)} }$ in QD and $\psi_{100}^{^{(1)} }$, $\psi
_{110}^{^{(1)} }$, $\psi_{11-1}^{^{(1)} }$ in SL at $B=40$~T.
 \label{tabl:table1}}
\vspace{1ex}
\begin{tabular}{|c||c|c|c|c|c|c|} \hline
& $\psi _{100}^{^{(0)} }$ & $\psi _{100}^{^{(1)} }$ & $\psi _{110}^{^{(0)} }$ & $\psi _{110}^{^{(1)} }$ & $\psi _{11-1}^{^{(0)} }$ & $\psi _{11-1}^{^{(1)} }$ \\ \hline\hline
$c_{10}$  & 0.9986 & 0.7018 & -- & -- & -- & -- \\ \hline $c_{11}$ & --
& -- & 0.9993 & 0.9151 & 0.9991 & 0.7949 \\ \hline $c_{12}$ &
0.0378 & 0.6888 & -- & -- & -- & -- \\ \hline $c_{13}$  & -- & -- &
0.0336 & 0.3969 & 0.0282 & 0.5923 \\ \hline $c_{14}$  & 0.0007 &
0.1801 & -- & -- & -- & -- \\ \hline $c_{15}$  & -- & -- & 0.0006 &
0.0705 & 0.0005 & 0.1306 \\ \hline $c_{16}$ & 0 & 0.0232 & -- & -- &
-- & -- \\ \hline $c_{17}$  & -- & -- & 0 & 0,0069 & 0 & 0.0144 \\
\hline $c_{20}$  & 0.0360 & 0.0044 & -- & -- & -- & --
\\ \hline $c_{21}$  & -- & -- & 0.0161 & 0.0044 & 0.0331 &
0.0087 \\ \hline $c_{22}$  & 0.0009 & 0.0014 & -- & -- & -- & --
\\ \hline $c_{23}$  & -- & -- & 0.0010 & 0.0007 & 0.0007 &
0.0022 \\ \hline $\sum \limits_{i} c_{i}^{2}  $ & 1.000 & 1.000 &
1.000 & 1.000 & 1.000 & 1.000 \\ \hline

\end{tabular}
\end{table}

\begin{figure}[!b]
\centerline{
\includegraphics[width=0.49\textwidth]{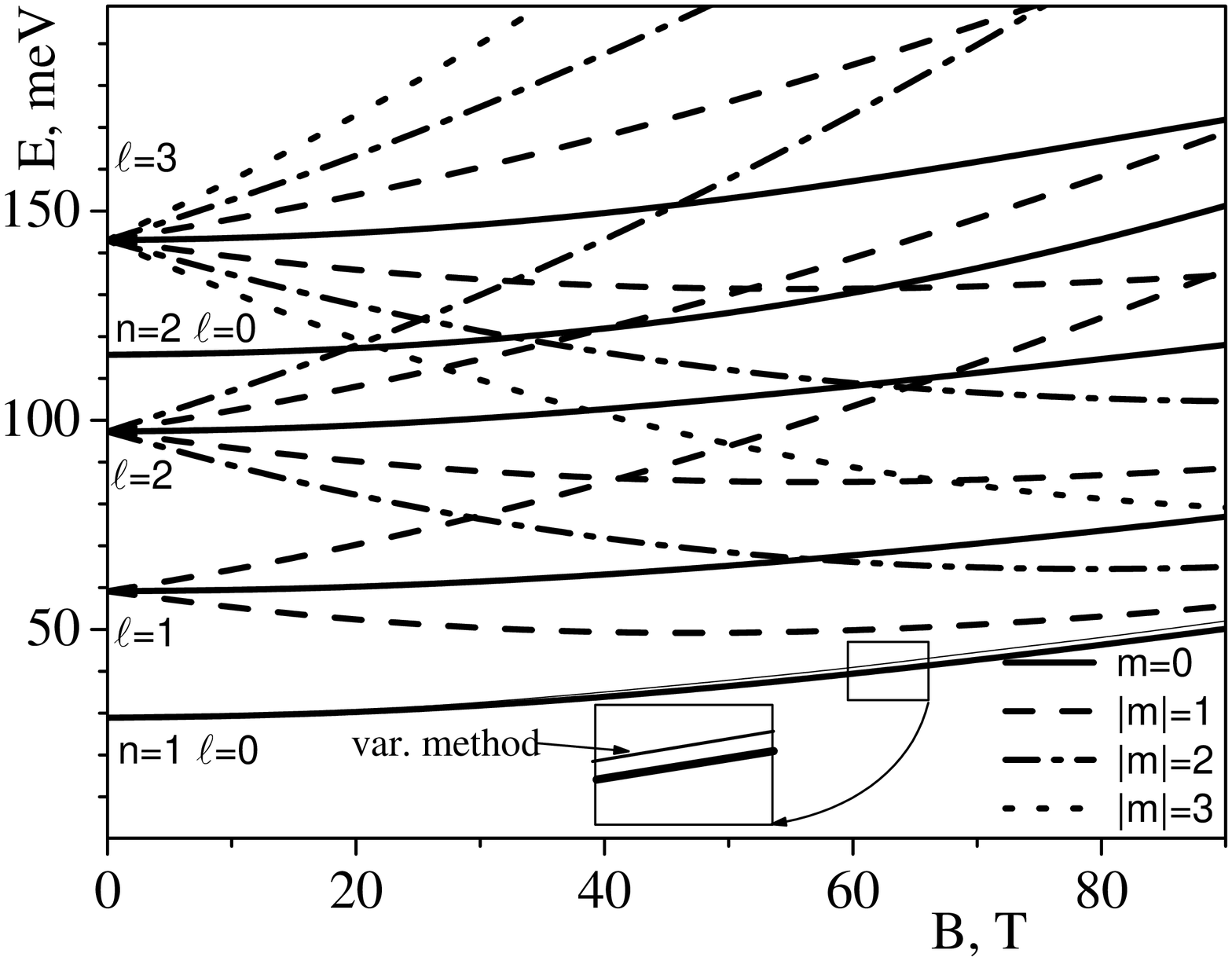}
\includegraphics[width=0.5\textwidth]{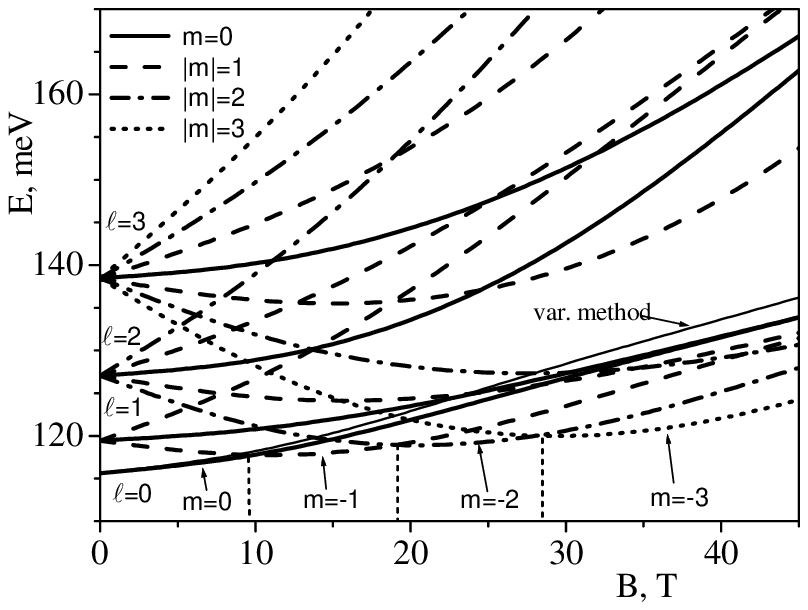}
}
\centerline{
\hspace{0.05\textwidth}(a) \hspace{0.45\textwidth} (b)
}
\caption{Electron energy spectrum as function of magnetic field
induction in QD with $r_0 = 10$~nm (a) and SL with $r_1 = 10$~nm,
$r_2 = 15$~nm (b).
\label{f1}}
\end{figure}

The dependencies of electron energy spectrum on the magnetic field
induction in CdSe QD and SL are presented in figure~\ref{f1}.

From figure~\ref{f1} one can see that the energy of $\psi_{100}$  state
calculated by a variational method correlates well to the one
obtained by the matrix method. Moreover, even in the case of
a strong magnetic field, the error for the electron energy in both
nano-structures does not exceed 3\%. Comparing the dependencies
shown in figures~\ref{f1}~(a) and \ref{f1}~(b), one can see that the magnetic field
produces a greater effect on the energy states of an electron located in SL than
on the energy states in QD. In both nano-structures, the degeneracy over the
magnetic quantum number is removed. The energies of the states
with $m \geqslant 0$ increase under the effect of the magnetic
field. For the states with $m < 0$, the non-monotonous dependence
of the energy on the magnetic field is caused by the linear and
quadratic terms contributed by the magnetic field into the
Hamiltonian (\ref{EQ.8}).

\begin{figure}[!t]
\centerline{ (a)
\includegraphics[width=0.6\textwidth]{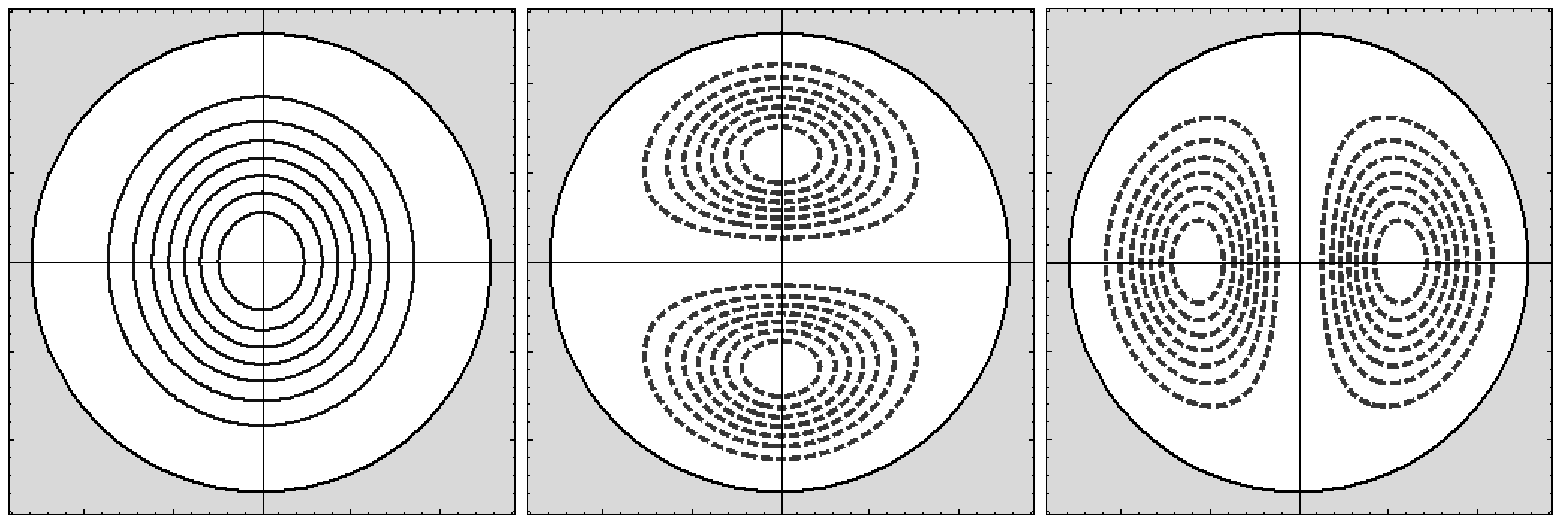}
}
\centerline{ (b)
\includegraphics[width=0.6\textwidth]{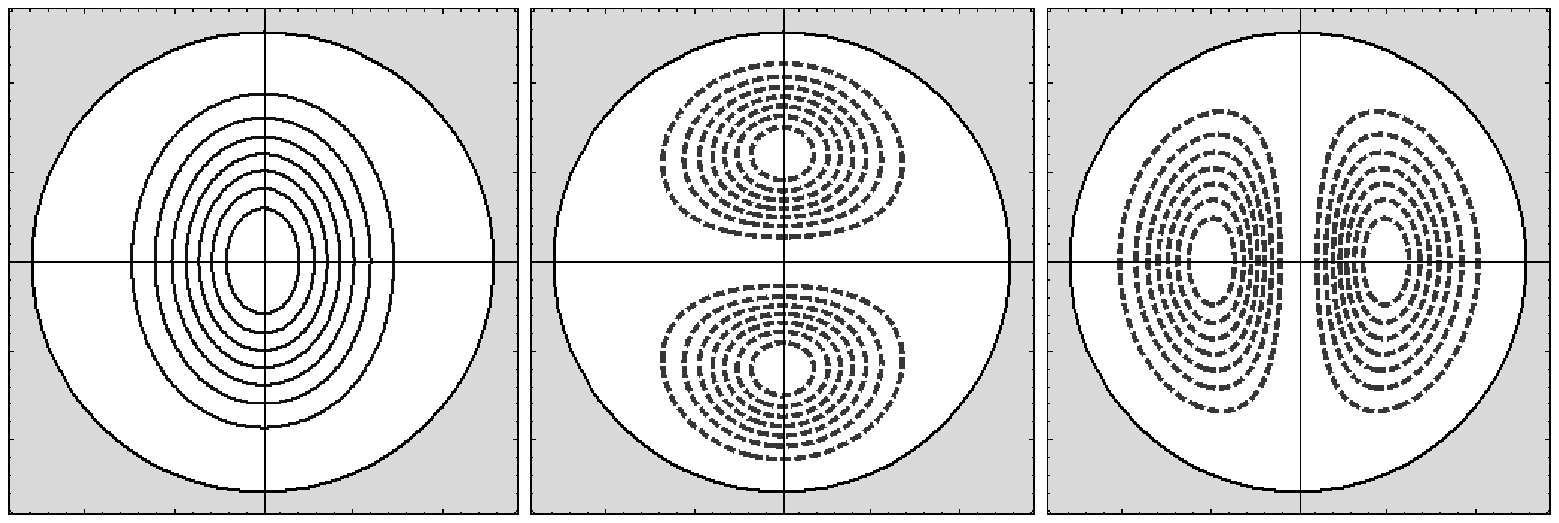}
}
\centering \caption{Distribution of probability density of
electron location in QD with  $r_0 = 10$~nm, at $B=40$~T (a),
$B=80$~T (b) for the quantum states with $\psi_{100}^{(0)}$,
$\psi_{110}^{(0)}$, $\psi_{11-1}^{(0)}$.
\label{f2}}
\end{figure}

\begin{figure}[!h]
\centerline{
(a)
\includegraphics[width=0.6\textwidth]{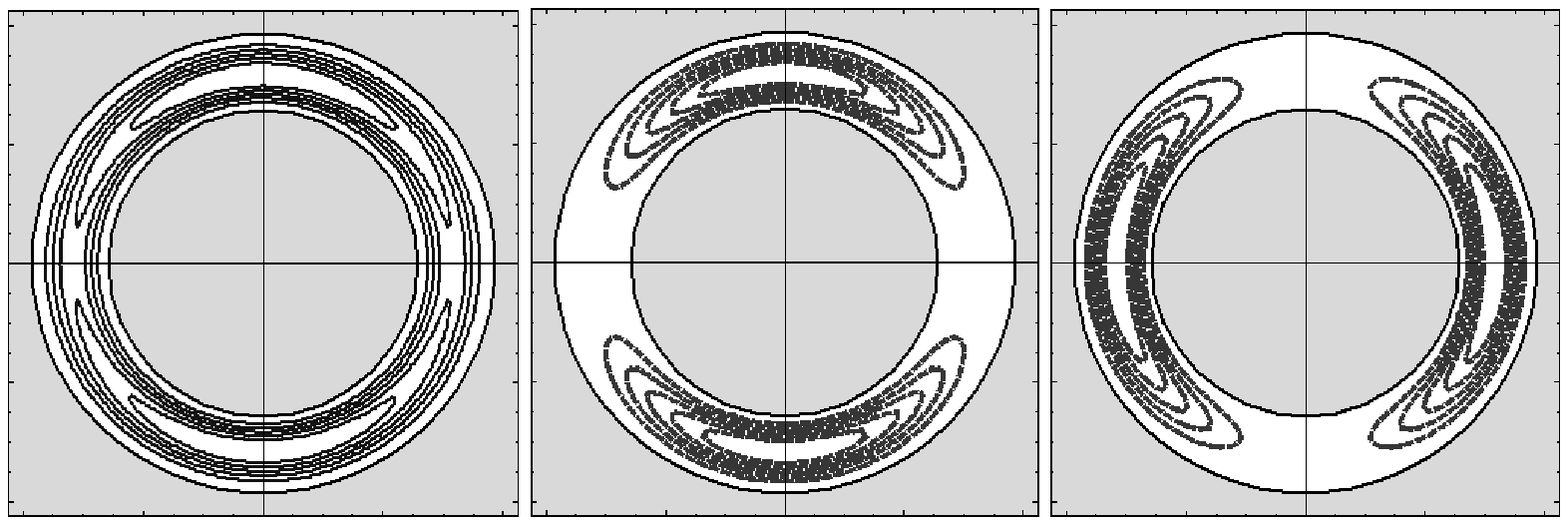}
}
\centerline{
(b)
\includegraphics[width=0.595\textwidth]{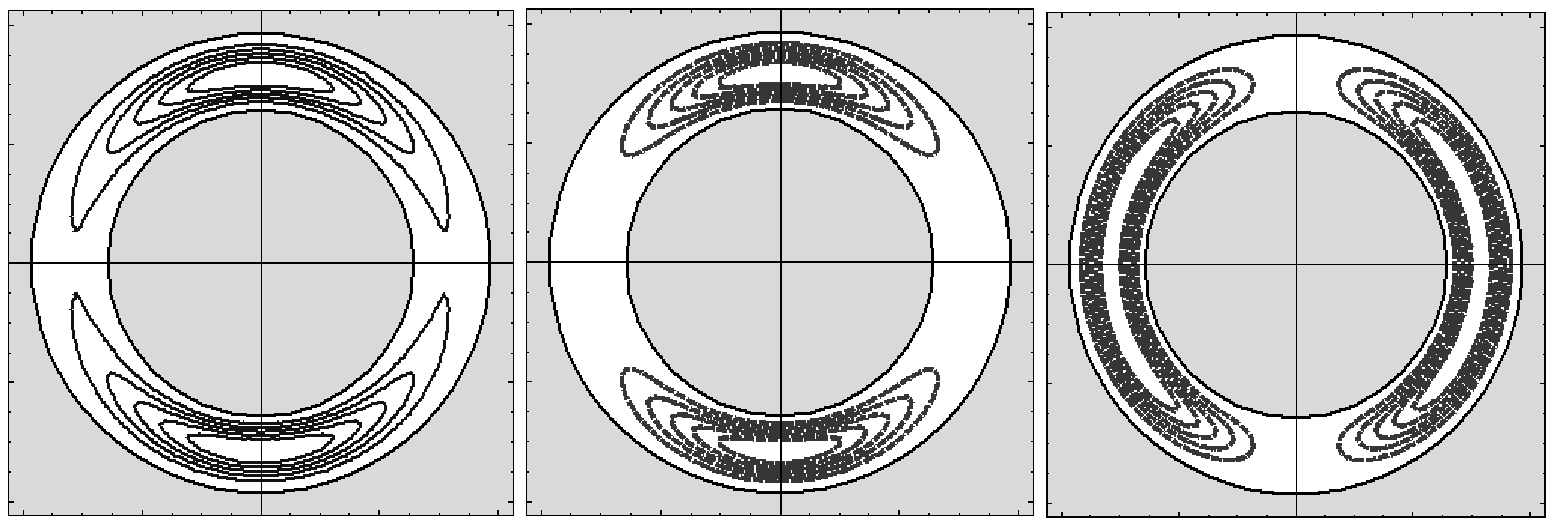}
}
\centerline{
(c)
\includegraphics[width=0.6\textwidth]{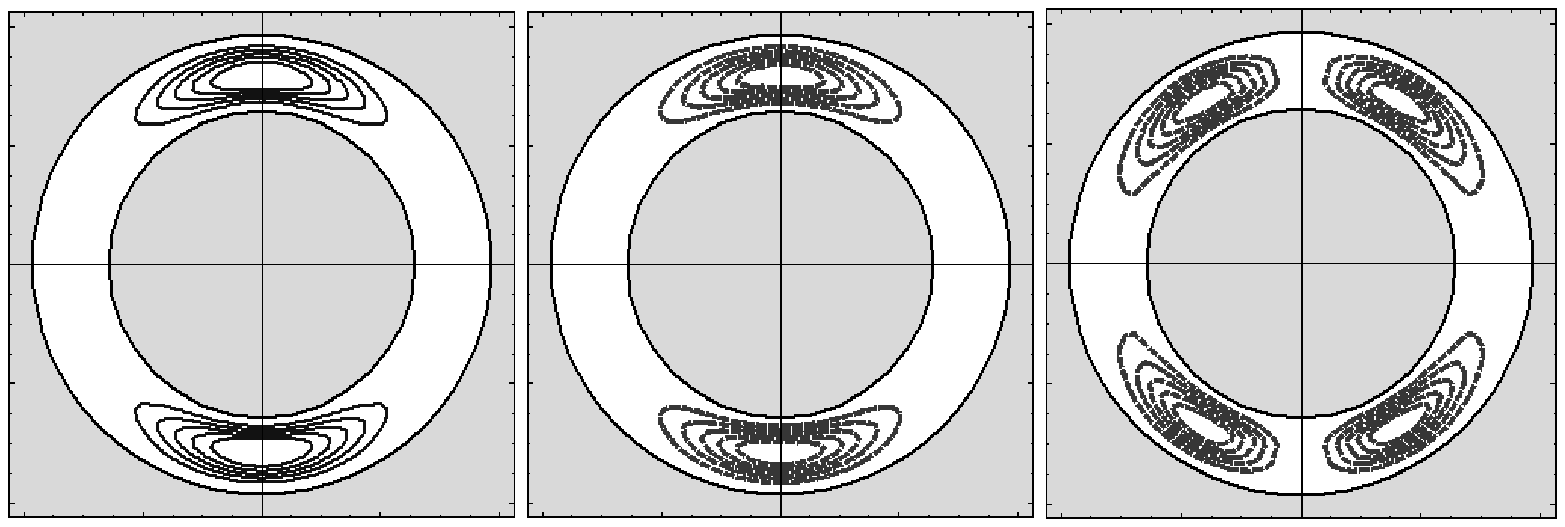}
}
 \centering \caption{Distribution of probability density of electron location
 in SL with  $r_1 = 10$~nm, $r_2 = 15$~nm at $B=10$~T (a), $B=20$~T (b) and $B=40$~T (c)
 for the quantum states with $\psi_{100}^{(1)}$,
$\psi_{110}^{(1)}$, $\psi_{11-1}^{(1)}$.
\label{f3}}
\end{figure}

\newpage

In the SL placed into a strong magnetic field, $1s$ and $1p$ states
with $m=0$ are degenerated, unlike in the QD. However, in
zero-dimensional systems, the electron ground state is always
non-degenerated. Consequently, when the magnetic field increases,
the lowest states with $m=0, -1, -2, \ldots$ successively play the role of
the ground state. The ground state of an electron  with a certain
value of a magnetic quantum number transforms into the state with
the other $m$ when the magnetic field intensity increases at an
equal magnitude. The distance between the points of such
a transition increases when SL radius becomes smaller. A similar
behavior of the electron ground state energy was theoretically
obtained and experimentally confirmed for the semiconductor
quantum rings \cite{22}. The oscillation of energies is known as
Aharonov Bohm effect.

 The distribution of probability density of
electron location in QD and SL in different quantum states is
presented in figures~\ref{f2}, \ref{f3}, respectively.

\begin{figure}[!h]
\centerline{
\includegraphics[width=0.8\textwidth]{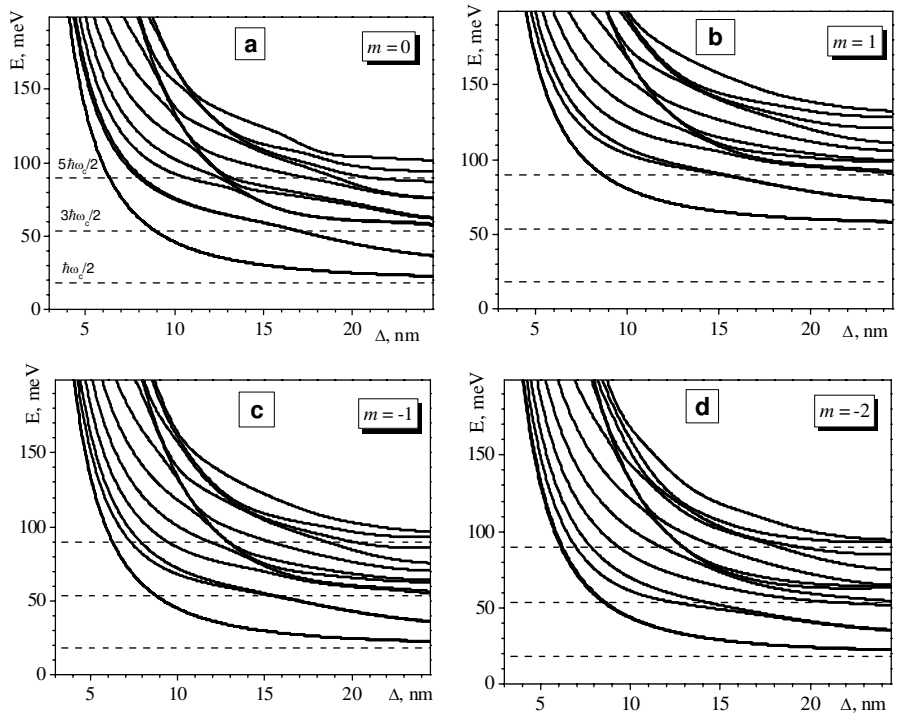}
}
\caption{Electron energy spectrum as function of $\Delta=r_2-r_1$ at
 $B=40$~T for $m=0$ (a), $m=1$ (b), $m=-1$ (c) and $m=-2$ (d). Landau levels in bulk
crystal are shown by dash lines.
\label{f4}}
\end{figure}

Figures~\ref{f2}, \ref{f3}, prove that the electron wave functions are deformed
due to the effect of the strong magnetic field. When its
induction increases, the angular probability increases near
$\theta=0, \pi $ and decreases near $\theta=\pi/2 $. Herein, in
the SL the wave function $\psi_{100}^{(1)}$ of $1s$ state, due to
the deformation, becomes similar to the $\psi_{110}^{(1)}$ one of
the excited $1p$ state. At $B=40$~T, these states become
indistinguishable both for the distribution of probability density
[figure~\ref{f3}~(c)] and the energy [figure~\ref{f1}~(b)].

Computer calculations prove that in the limit case when the inner
radius of SL $r_1$ diminishes at $r_2=\textrm{const}$ and at the constant
magnetic field induction, the electron energy spectrum coincides
with the one for the QD with $r_0= r_2$.

In limit cases, when the QD radius or SL thickness increases at
$B=\mathrm{const}$ due to a decrease of the quantum confining effect, the
electron energy levels should coincide with Landau levels which
are typical of a  bulk crystal placed into the magnetic field. The
process of the formation of Landau levels for the quantum states
with $m=0,1,-1$ in the SL is presented in figure~\ref{f4}.

Figure~\ref{f4} proves that the quantum confining effect diminishes when
the sizes of the structure increase. This process is accompanied
by a decrease of all energy levels and by the formation of Landau
levels. For example, at $m=0$, the lowest Landau level is formed by
the set of levels with $n=1$, the next one with $n=2$ and so on. A
complicated dependence of an energy spectrum is observed during their
formation due to the anti-crossing effect.

\section{Summary}

We studied the electron energy spectrum in QD and SL under the
effect of a magnetic field. The problem is solved using the
variational method and the method of electron wave function
expansion over the set of eigenfunctions being the exact solutions
of Schr\"odinger equation for the same structures without the
magnetic field. The results obtained within the both methods are
in good agreement. The variational method describes the lowest electron state with $m=0$. However, the ground state in SL is formed by the states with $m=0, -1, -2, \ldots$ consequently with the growth of the magnetic field induction.
It is shown that the major contribution  into
the expansion of a wave function of an arbitrary electron state even in
the strong magnetic field is performed by a few neighboring (over the
energy) quantum states which are the exact solutions of
Schr\"odinger equation for the electron when there is no magnetic field.

The wave functions of the electron in QD and SL are deformed under
the effect of a magnetic field. The degeneration of an energy
spectrum with respect to a magnetic quantum number is removed. The
electron energies for the states with positive and negative values
of a magnetic quantum number differently depend on the magnetic field
induction: for the states with $m \geqslant 0$, the energy monotonously
increases and for $m<0$, the energy decreases at first and then,
only when the magnetic field becomes strong enough, enhances.

It is proven that the effect of a magnetic field on the electron
energy spectrum in SL is stronger than that in QD. Moreover, it is
shown that in SL driven by the strong magnetic field, the neighboring
states with the same $m$ become degenerated. For example, in the
studied structure, $1s$ and $1p$ states are degenerated at $B>30$~T.
The degeneration of higher energy states takes place at a stronger
magnetic field.

The validity of the obtained results is confirmed by the limit cases:
when the quantum confining effect reduces at a constant magnetic
field induction, the energy levels rebuild, saturate and form the
respective Landau levels. When the inner radius of SL ($r_1$)
decreases at $r_2=\textrm{const}$, the electron energy spectrum becomes the
same as in the QD having the  radius $r_2$.

The results of the investigation make it possible to estimate the energies
and the most probable place of electron location in QDQW.
Different dependencies of electron energies in QD and SL on
the magnetic field induction lead to their anticrossing in QDQW. This
feature permits to change the quasiparticle location by a magnetic
field for multilayered systems with penetrable interfaces.

\ukrainianpart

\title{Вплив магнітного поля на спектр електрона у сферичних наноструктурах}

\author{В.А. Головацький, О.М. Войцехівська, І.Б. Бернік}
\address{
Чернівецький національний університет ім. Юрія Федьковича, вул.
Коцюбинського, 2,  58012 Чернівці, Україна}

\makeukrtitle

\begin{abstract}
\tolerance=3000%
Досліджено вплив постійного магнітного поля на енергетичний спектр та хвильові
функції електрона в сферичних наноструктурах: простій квантовій точці та
сферичній плівці. Показано, що під дією магнітного поля знімається виродження енергетичного
спектру
за магнітним квантовим числом. Збільшення індукції магнітного поля приводить
до монотонного зростання енергії станів з магнітним квантовим числом $m\geqslant 0$
та немонотонної залежності енергії станів з $m<0$.
Досліджено вплив магнітного поля на хвильовi функцiї основного та декiлькох
збуджених станів електрона. У сферичній плівці зі збільшенням індукції
магнітного поля відбувається виродження $1s$ та $1p$ станів. При збільшенні
розмірів наносистеми чи збільшенні індукції магнітного поля відбувається
формування серій рівнів Ландау, що характерні для масивного кристалу.

\keywords спектр електрона, квантова точка, сферична плівка,
магнітне поле

\end{abstract}


\begin{thebibliography}{99} %


\bibitem{1} Eychmuller A., Mews A., Weller H., Chem. Phys. Lett., 1993, \textbf{208},
59; \doi{10.1016/0009-2614(93)80076-2}.

\bibitem{2} Dorfs D., Eychmuller A., Z. Phys. Chem., 2006, \textbf{220},
1539; \doi{10.1524/zpch.2006.220.12.1539}.

\bibitem{3} Little R., El-Sayed M., Bryant G., Burke S., Chem. Phys., 2001, \textbf{114}, 1813.

\bibitem{4} Frasco M.F., Chaniotakis  N., Sensors, 2009, \textbf{9},
7266; \doi{10.3390/s90907266}.

\bibitem{5} Liu Y.S., Sun Y., Vernier P.T., Liang C. H., Chong S.Y.,
Gundersen  M.A., J. Phys. Chem. C. Nanomater Interfaces, 2007,
\textbf{111}, 2872; \doi{10.1021/jp0654718}.

\bibitem{6} Demir H., Nizamoglu S., Mutlugun E., Ozel T., Sapra S.,
Gaponik N., Eychmuller A., Nanotechnology, 2008, \textbf{19},
335203; \doi{10.1063/1.2898892}.

\bibitem{7} SalmanOgli A., Rostami A., J. Nanopart. Res., 2011, \textbf{13}, 1197; \doi{10.1007/s11051-010-0112-2}.

\bibitem{8} Tkach N., Voitsekhovska O., Holovatsky V., Mihalyova M.,
Izvestiya vuzov. Fizika, 1998, \textbf{12}, 58 (in Russian) [Russ. Phys. J., 1998, \textbf{41}, 1229; \doi{10.1007/BF02514561}].

\bibitem{9} Holovatsky  V., J. Phys. Stud., 1998, \textbf{2}, 583 (in Ukrainian).

\bibitem{10} Tkach M., Holovatsky V., Voitsekhivska O., Mikhalyova M.,
Electrochemical Society Proceedings, 1998, \textbf{25}, 316.

\bibitem{11} Tkach  M., Holovatsky V., Voitsekhivska O., Fiz. Teh. Pol., 2000, \textbf{34}, 602 (in Russian) [Semiconductors, 2000, \textbf{34}, 583; \doi{10.1134/1.1188032}].

\bibitem{12} Holovatsky V., Makhanets O., Voitsekhivska O., Physica E, 2009, \textbf{41}, 1522; \doi{10.1016/j.physe.2009.04.027}.

\bibitem{13} Boichuk V.I., Bilynskyi I.V., Leshko R.Ya., Voronyak L.Ya.,
Ukr. J. Phys., 2009, \textbf{54}, 1021.

\bibitem{14} Tas H.,  Sahin M., J. Appl. Phys., 2012, \textbf{111},
083702; \doi{10.1063/1.4751483}.

\bibitem{15} Wu S., Wan L., J. Appl. Phys., 2012, \textbf{111},
063711; \doi{10.1063/1.3695454}.

\bibitem{16} Rahmani K., Zorkani I., M. J. Condensed Matter., 2009, \textbf{11}, 35.

\bibitem{17} Chakraborty T., Apalkov V., Physica E, 2003, \textbf{16}, 253; \doi{10.1016/S1386-9477(02)00674-4}.

\bibitem{18} Xiao Z., J. Appl. Phys., 1999, \textbf{86}, 4509; \doi{10.1063/1.371394}.

\bibitem{19} Planelles J., Diaz J., Climente J., Jaskolski W.,  Phys. Rev. B, 2002, \textbf{65}, 245302; \doi{10.1103/PhysRevB.65.245302}.

\bibitem{20} Battaglia D., Blackman B., Peng X., J. Am. Chem. Soc., 2005, \textbf{127},
10889-10897; \doi{10.1021/ja0437297}.

\bibitem{21} Jiang  H., Phys. Rev. B., 1987, \textbf{35}, 9287;
\doi{10.1103/PhysRevB.35.9287}.

\bibitem{22} Lorke A., Luyken R., Govorov A., Kotthaus J.J., Garcia M.,
Petroff P., Phys. Rev. B,  2000, \textbf{84}, 2223;
\doi{10.1103/PhysRevLett.84.2223}.

\end{thebibliography}
\end{document}